\documentclass[aps,pre,twocolumn,superscriptaddress,showpacs]{revtex4-1}
\usepackage{amsfonts}
\usepackage{epsfig}
\usepackage{amsmath}
\usepackage{times}
\usepackage{color}

\begin{document}

\title{GOE statistics in graphene billiards with the shape of classically integrable billiards}

\author{Pei Yu}
\affiliation{School of Physical Science and Technology, and Key
Laboratory for Magnetism and Magnetic Materials of MOE, Lanzhou University,
Lanzhou, Gansu 730000, China}

\author{Zi-Yuan Li}
\affiliation{School of Physical Science and Technology, and Key
Laboratory for Magnetism and Magnetic Materials of MOE, Lanzhou University,
Lanzhou, Gansu 730000, China}

\author{Hong-Ya Xu}
\affiliation{School of Electrical, Computer, and Energy Engineering,
Arizona State University, Tempe, Arizona 85287, USA}

\author{Liang Huang}\email{huangl@lzu.edu.cn}
\affiliation{School of Physical Science and Technology, and Key
Laboratory for Magnetism and Magnetic Materials of MOE, Lanzhou University,
Lanzhou, Gansu 730000, China}

\author{Barbara Dietz}
\affiliation{School of Physical Science and Technology, and Key
Laboratory for Magnetism and Magnetic Materials of MOE, Lanzhou University,
Lanzhou, Gansu 730000, China}

\author{Celso Grebogi}
\affiliation{Institute for Complex Systems and Mathematical
Biology, King's College, University of Aberdeen, Aberdeen AB24
3UE, UK}

\author{Ying-Cheng Lai}
\affiliation{School of Electrical, Computer, and Energy Engineering,
Arizona State University, Tempe, Arizona 85287, USA}
\affiliation{Institute for Complex Systems and Mathematical
Biology, King's College, University of Aberdeen, Aberdeen AB24
3UE, UK}
\affiliation{Department of Physics, Arizona State University, Tempe,
Arizona 85287, USA}

\date{\today}

\begin{abstract}

A crucial result in quantum chaos, which has been established for a long
time, is that the spectral properties of classically integrable systems
generically are described by Poisson statistics whereas those of time-reversal symmetric, classically chaotic systems coincide with those of random matrices from the Gaussian orthogonal ensemble (GOE). Does this result hold for two-dimensional Dirac material systems? To address this fundamental question, we investigate the
spectral properties in a representative class of graphene billiards with shapes of classically integrable circular-sector billiards. Naively one may
expect to observe Poisson statistics, which is indeed true for energies
close to the band edges where the quasiparticle obeys the Schr\"{o}dinger
equation. However, for energies near the Dirac point, where the quasiparticles
behave like massless Dirac fermions, Poisson statistics is extremely rare
in the sense that it emerges only under quite strict symmetry constraints
on the straight boundary parts of the sector. An arbitrarily small amount of
imperfection of the boundary results in GOE statistics. This implies that, for
circular sector confinements with arbitrary angle, the spectral
properties will generically be GOE. These results are corroborated by
extensive numerical computation. Furthermore, we provide a physical understanding for our results.

\end{abstract}

\pacs{05.45.Mt, 81.05.ue, 73.23.-b}
\maketitle

\section{Introduction} \label{sec:intro}

A fundamental result in quantum chaos, a field that studies the quantum
signatures of classical chaos, is that distinct properties of the classical dynamics lead to characteristically different fluctuation properties in the energy spectra of the corresponding quantum system. In particular, for classically integrable systems, the energy levels behave like random numbers from a Poisson process, whereas the spectral properties of generic, classically chaotic systems coincide with those of the eigenvalues of random matrices from the  Gaussian orthogonal ensemble (GOE), if time-reversal invariance is preserved~\cite{BGS:1984,Haake:book,Stockmann:book}. When time-reversal invariance is violated, the statistics is described by the Gaussian unitary ensembles (GUE). We note, that in nonrelativistic quantum chaotic systems, a magnetic field induces time-reversal symmetry breaking and changes the level statistics from GOE to GUE~\cite{SAOO:1995,SSSKH:1995,LSS:1996,KR:1997,DGHRRS:2000}. Although the above correspondences may be violated for certain nongeneric cases, they are generally expected to hold for typical nonrelativistic quantum systems and even for relativistic quasiparticles in two-dimensional systems governed by the Dirac equation such as graphene flakes, also called graphene billiards~\cite{PSKYHNG:2008,Wurmetal:2009,HLG:2010,HLG:2011,Dietz2015,HXLG:2014}.

Berry and Mondragon found in their seminal work~\cite{BM:1987} that for relativistic neutrino billiards, i.e., massless spin-1/2 particles governed by the two-dimensional Dirac equation and confined to a bounded region, the spectral properties follow the Poisson statistics, if the shape of the confinement corresponds to a billiard with classically integrable dynamics, like a circular billiard. Here, a billiard is a bounded two-dimensional domain inside which a pointlike particle moves freely and is reflected specularly at the walls. On the contrary, when the shape of the neutrino billiard corresponds to that of a classically chaotic billiard, the spectral properties agree with those of random matrices from the GUE, even in the absence of a magnetic field. This feature is attributed to the time-reversal invariance violation induced by the mass confinement.

\begin{figure}[h]
\begin{center}
\epsfig{file=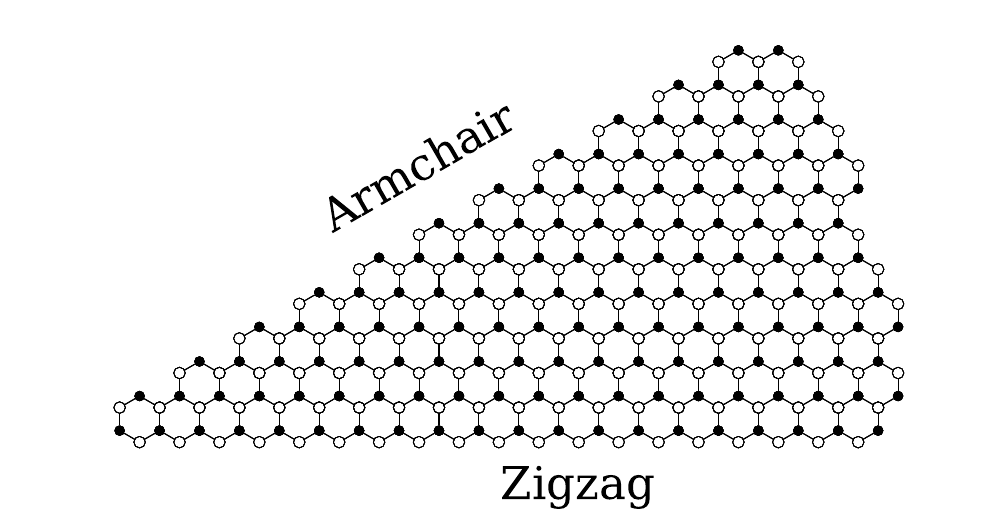,width=\linewidth}
\end{center}
\caption{
Schematic view of a graphene billiard with the shape of a circular sector. The hexagonal lattice is formed by two independent triangular sublattices marked by filled and empty circles, respectively. Similar to the circular part of the sector, the boundaries of the graphene billiards considered in the present article are formed by pieces of zigzag and armchair edges.
} \label{fig0}
\end{figure}

Since the discovery of graphene~\cite{Novoselovetal:2004,Bergeretal:2004,
Novoselovetal:2005,ZTSK:2005,Netoetal:2009,Peres:2010,SAHR:2011}, the
fluctuation properties in the energy spectra of graphene billiards, i.e., finite-size graphene sheets formed by a single-layer of atoms arranged on a honeycomb lattice,
have been investigated intensively. Figure~\ref{fig0} shows an example for a graphene billiard with the shape of a circular sector.
In view of the result of Berry and Mondragon on chaotic neutrino billiards, the spectral properties of chaotic graphene billiards were expected to follow GUE statistics (see Ref.~\cite{HXLG:2014} and references therein). However, extensive simulations using different chaotic graphene billiards clearly indicated GOE statistics~\cite{Wurmetal:2009,HLG:2010,HLG:2011}. The physical mechanism for the GOE rather than GUE statistics has its origin in the hexagonal lattice structure of graphene which is composed of two independent triangular sublattices. As a consequence, the conduction and the valence band exhibit conically shaped valleys that touch each other at the six corners of the hexagonal first Brillouin zone, which, like the graphene lattice, is composed of two independent triangles, each corresponding to one of the two independent Dirac points. In each of the independent valleys the electron excitations are governed by a two-dimensional Dirac equation for massless spin-1/2 quasiparticles. The components of the associated pseudospin are related to the wave function amplitudes on, respectively, one of the triangular sublattices of the honeycomb lattice~\cite{Beenakker2008,Netoetal:2009}.

The scattering at the boundaries of a finite-size graphene sheet leads to a mixing of the valleys. Accordingly, there the Dirac equations are coupled and the hexagonal lattice is described by a four-dimensional Dirac equation which preserves the time-reversal invariance. Naturally, it was speculated that a reduced degree of valley mixing might cause the level statistics to deviate from GOE toward GUE. In order to reduce the mixing, a smoothly varying mass potential was introduced in Ref.~\cite{Wurmetal:2009} in the vicinity of the domain boundary in order to diminish the intervalley scattering. However, in general, the spectral statistics tended to GOE statistics irrespective of the residual scattering, which hence seems to be non-negligible. Consequently, GUE statistics may only be observed in graphene billiards with zigzag edges formed by atoms from the same sublattice like in an equilateral triangle~\cite{Rycerz:2012}. In this case, introducing a smoothly varying disorder potential will induce a transition from Poisson statistics to GOE statistics, if all three sides are formed by zigzag edges, and to GUE statistics if the sheet is terminated by armchair edges~\cite{Rycerz:2012}. Figure~\ref{fig0} shows for illustration the structure of a zigzag and an armchair edge. However, in the latter case, weak disorder can turn the statistics back to GOE in larger systems. It was subsequently shown~\cite{Rycerz:2013} that, if a zigzag triangular nanoflake is deformed so that it does not possess any geometric
symmetry, a strain-induced gauge field can effectively break the time-reversal symmetry, leading to GUE statistics. Furthermore, the effects of random disorder
and edge roughness were studied in Refs.~\cite{LSB:2009,AE:2009} with the result
that increasing disorder or edge roughness is accompanied by
a transition from Poisson to GOE statistics.

Current understanding of the energy level statistics of graphene billiards
can be summarized as follows. Without magnetic field, if the shape of the bounded
domain leads to a classically integrable dynamics, the spectral fluctuations
coincide with that of Poissonian random numbers~\cite{LSB:2009,AE:2009}.
If the shape of the confinement coincides with that of a billiard with classically chaotic
dynamics, the statistics is described by the GOE~\cite{Wurmetal:2009,HLG:2010,HLG:2011}.
The main result of the present article is the uncovering of counterintuitive
features of the spectral statistics: finite-size graphene sheets with the shapes of
certain classes of classically integrable billiards generically exhibit GOE
statistics. We present extensive numerical results and physical reasonings
to establish our finding.

\section{Model} \label{sec:model}

We investigate the fluctuation properties in the eigenvalue spectra of finite-size graphene sheets, so-called graphene billiards with shapes of circular sectors of angles $\pi/n$, where $n$ is an integer. The corresponding classical billiards exhibit an integrable dynamics. The confined billiard domains were cut from a perfect graphene sheet. For most cases, the orientation was chosen such that one of the straight boundaries of the sector was formed by a zigzag edge.

Qualitative~\cite{Wallace:1947} and also some quantitative insight~\cite{Reich2002} into the band structure of infinitely extended graphene has been obtained using a tight-binding model. Its application is based on the assumption that interactions of the graphene $p_z$ orbitals are nonnegligible only for nearest, second-nearest and third-nearest neighbors. The same model may be used to determine the energy levels in finite-size graphene sheets. The dimension of the associated tight-binding Hamiltonian coincides with the number of atoms forming the graphene sheet, with each diagonal element corresponding to one atom. This corresponds to imposing Dirichlet boundary conditions along the first row of atoms outside the boundary of the graphene billiard. Setting the on-site energies of the isolated atoms equal to zero and taking into account only couplings between neighboring atoms, the tight-binding Hamiltonian is given by $\widehat{H}=\sum -t |i\rangle\langle j|$, where the summation is over all pairs of nearest-neighbor atoms, and $t\approx 2.8$eV is the corresponding hopping energy~\cite{Novoselovetal:2004,Bergeretal:2004,
Novoselovetal:2005,ZTSK:2005,Netoetal:2009,Peres:2010,SAHR:2011}.
Diagonalizing the Hamiltonian matrix yields the energy eigenvalues and
eigenfunctions of the system. We diagonalized matrices corresponding to lattice sizes of up to $2\times 10^5$ atoms. For the presentation of the characteristic eigenfunction patterns it was sufficient to consider lattices consisting of $\approx 5\times 10^4$ atoms.

\begin{table}[h]
\caption{(Color online) Graphene sector billiards investigated in this paper, where ``N'' denotes the non-perfect cases where a few atoms at the corner or one row of atoms at one of the straight boundary was missing, i.e., their shapes slightly deviated from that of a sector billiard. The letters ``Z'' and ``A'' stand for zigzag and armchair edges along the straight boundaries (see Fig.~\ref{fig0}), respectively. The size of the graphene sheets is specified by the number of lattice sites.}
\label{tab:sectors}
\begin{ruledtabular}
\begin{tabular}{lccccc}
\textrm{Angle}&
\multicolumn{1}{c}{$15^{\circ}$}&
\multicolumn{1}{c}{$45^{\circ}$}&
\multicolumn{1}{c}{$30^{\circ}$}&
$30^{\circ}$\textrm{N}\\
\colrule
 \textrm{Size} & 226993 & 244664 & 220862 &221944\\
\colrule
  & & & & \\
\colrule
 \textrm{Angle} & $90^{\circ}$ & $60^{\circ}$\textrm{Z} & $60^{\circ}$\textrm{A} & $60^{\circ}$\textrm{AN}\\
\colrule
 \textrm{Size} & 217615 & 226042 & 227254 & 226315\\
\end{tabular}
\end{ruledtabular}
\end{table}

Table~\ref{tab:sectors} lists the lattice structures considered in the present article.
To be more specific, we considered altogether 8 sectors with varying edge structures along the straight boundaries and angles $\pi /n$ with $n=2,3,4,6,12$, corresponding to $15^{\circ}, 30^{\circ}, 45^{\circ}, 60^{\circ}, 90^{\circ}$, respectively, whereas the curved boundary was obtained by fitting the hexagonal lattice into the corresponding sector billiard. For the $15^{\circ}$ and $45^{\circ}$ sectors, one straight boundary was terminated by a zigzag edge, while the other one consisted of a mixture of zigzag and armchair segments. For the $30^{\circ}$ and $90^{\circ}$ domains, a perfect cut resulted in a zigzag edge for one straight boundary and an armchair edge for the other one, so there was no reflection symmetry. For the $30^{\circ}$ sector, besides the perfect-cut case, also the non-perfect cut case was studied, the only difference being one additional layer of atoms along the armchair edge. A $60^{\circ}$-sector graphene billiard can be cut out of a hexagonal lattice such that both straight boundaries are zigzag or armchair edges. Accordingly, these graphene billiards can possess an exact reflection symmetry. In addition, we
considered a $60^{\circ}$ lattice structure in which the reflection symmetry was broken by removing one atomic layer along one of the straight boundaries.

Taking into account only the nearest-neighbor hopping term in the Hamiltonian yields eigenstates that are symmetric under the $E\rightarrow -E$ and the $[\Psi_A, \Psi_B] \rightarrow [\Psi_A, -\Psi_B]$ operations~\cite{BF:2006}. Accordingly, it is sufficient to consider only states corresponding to positive energy values, $0 \alt E/t \le 3$. The quasiparticles in graphene exhibit distinct behaviors in different energy ranges~\cite{Dietz2015}. In particular,
in the lower energy range close to the Dirac point, $E/t \in [0.02,0.2]$, the quasiparticles are described by the two-dimensional Dirac equation for massless Dirac fermions with little trigonal warping so that the corresponding motion is isotropic. Note, that by restricting to $E\gtrsim 0.02$ we discard eigenvalues corresponding to edge states~\cite{HLG:2011} present in the vicinity of the Dirac point. For higher energies, trigonal warping becomes important, especially close to the van Hove singularities at $E/t=\pm 1$. There, the motion of the quasiparticles is limited to the six directions parallel to zigzag edges within the graphene billiard. This becomes visible in the intensity distributions of the wavefunctions, which exhibit a striped structures along these directions~\cite{DKMR:2016}. Close to the band edges $E_{edge}=\pm 3t$, the dispersion relation becomes parabolic. There, the quasiparticles are governed by the nonrelativistic Schr\"{o}dinger equation. Thus we consider a quantum billiard with the same confinement as the graphene sector billiard.

The eigenstates of a quantum billiard in a sector domain~\cite{Dietz1995} of unit radius and arbitrary
angle $\alpha$, i.e., of a quantum particle trapped in an infinitely high potential well with the shape of a sector, are obtained by solving the Schr\"{o}dinger equation with Dirichlet boundary conditions along the border of the domain. They are given in terms of the Bessel functions, $\Psi_{mn}\propto\sin(m\pi\varphi/\alpha)J_{m\pi/\alpha}(k_{mn}\rho)$, where $\varphi$ and $\rho$ are the angular and radial polar coordinates, respectively, and the indices are integers, $m,n=1,2,3,\cdots$. The eigenwavevectors $k_{mn}>0$ are obtained from the zeros of the Bessel functions along the boundary, i.e., for $\rho =1$, $J_{m\pi/\alpha}(k_{mn})=0$, yielding the eigenenergies $E_{QB}=\hbar^2 k_{mn}^2/(2m)$, where $m$ is the mass and $\hbar$ is the reduced Planck constant.
For large $x$, we have $J_{m\pi/\alpha}(x) \sim \sqrt{2/(\pi x)}
\cos{(x- (m\pi/\alpha)\pi/2 -\pi/4)}$ so that asymptotically the zeros for
different $m$ are shifted by integer multiples of $\pi^2/2\alpha$ with respect to each other, similar to the situation discussed in Ref.~\cite{CCG:1985}, where the spectral properties were shown to be described by the Poisson statistics. Accordingly, the intuitive expectation is that the eigenvalues of sector-shaped graphene billiards exhibit a similar behavior regardless of the value of $\alpha$. However, as we shall demonstrate in the sequel, generally the results may be contrary to this speculation depending on the energy range under consideration.

\section{Results} \label{sec:results}
\begin{figure}[h]
\begin{center}
\epsfig{file=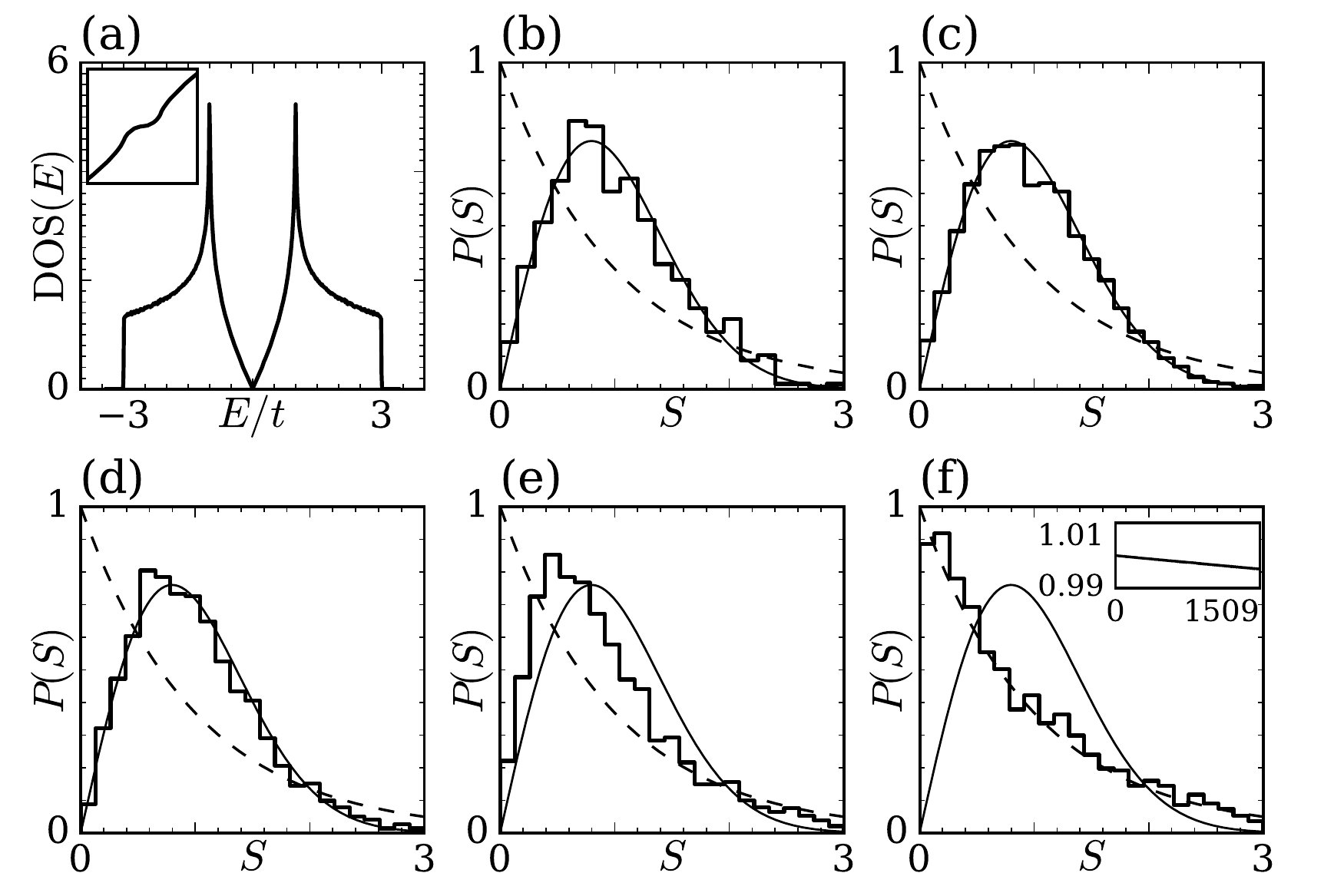,width=\linewidth}
\end{center}
\caption{Spectral properties of a graphene billiard with the shape of a $15^{\circ}$ sector.
(a) The density of states. Inset: integrated density of states. (b-f)
Nearest-neighbor spacing distributions in the energy ranges:
(b) $E/t \in [0.02,0.2]$, 837 levels; (c) $E/t \in [0.5,0.6]$, 2575 levels;
(d) $E/t \in [2.0,2.1]$, 3804 levels; (e) $E/t \in [2.6,2.7]$, 3321 levels;
(f) $E/t \in [2.95,3]$, 1509 levels. The histograms show the
numerical results, the dashed and solid lines exhibit
the Poisson and GOE distributions, respectively.
Inset in panel (f): normalized ratio $(E-E_{edge})/E_{QB}$ for the first 1509 eigenenergies, where $E_{QB}$ is the eigenenergy of the corresponding quantum billiards.}
\label{fig:15degree1}
\end{figure}

\begin{figure}[h]
\begin{center}
\epsfig{file=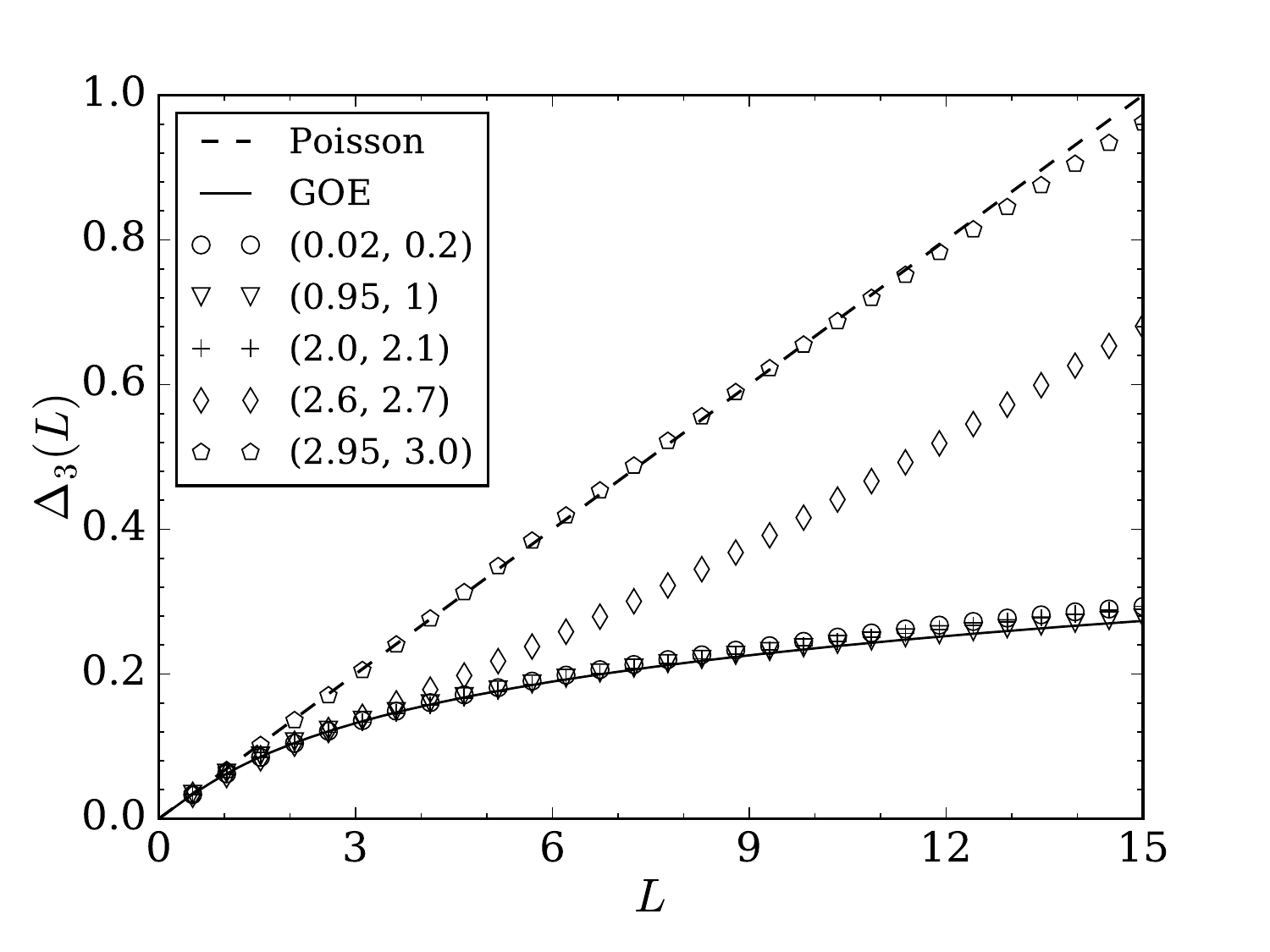,width=\linewidth}
\end{center}
\caption{$\Delta_3$ statistics of the $15^{\circ}$ sector. The energy ranges are the same as in Fig.~\ref{fig:15degree1}. The dashed and solid lines exhibit the Poisson and GOE results, respectively. }
\label{fig:15degree2}
\end{figure}

\begin{figure}[h]
\begin{center}
\epsfig{file=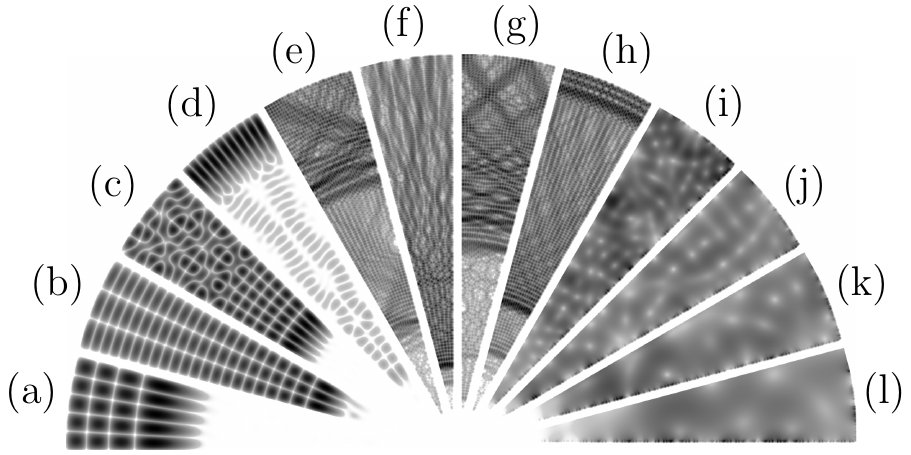,width=\linewidth}
\end{center}
\caption{Intensity distributions in the $15^{\circ}$-sector graphene billiard.
Panels (a-l) correspond to energy values 2.991, 2.981, 2.976,
2.974, 2.667, 2.664, 2.663, 2.661, 0.195, 0.140, 0.100, and 0.063,
respectively. In panels (a-d), the energies are close to the band edge,
corresponding to eigenstates with ($m=5$,$n=4$), ($m=3$,$n=22$),
($m=5$,$n=18$), and ($m=11$,$n=1$) in the quantum billiard of corresponding shape, respectively. Panels (e-h) show results for the transitional
region around $E/t \sim 2.66$. In panels (i-l) the energies are close to the Dirac point.}
\label{fig:15patterns}
\end{figure}

\begin{figure}[h]
\begin{center}
\epsfig{file=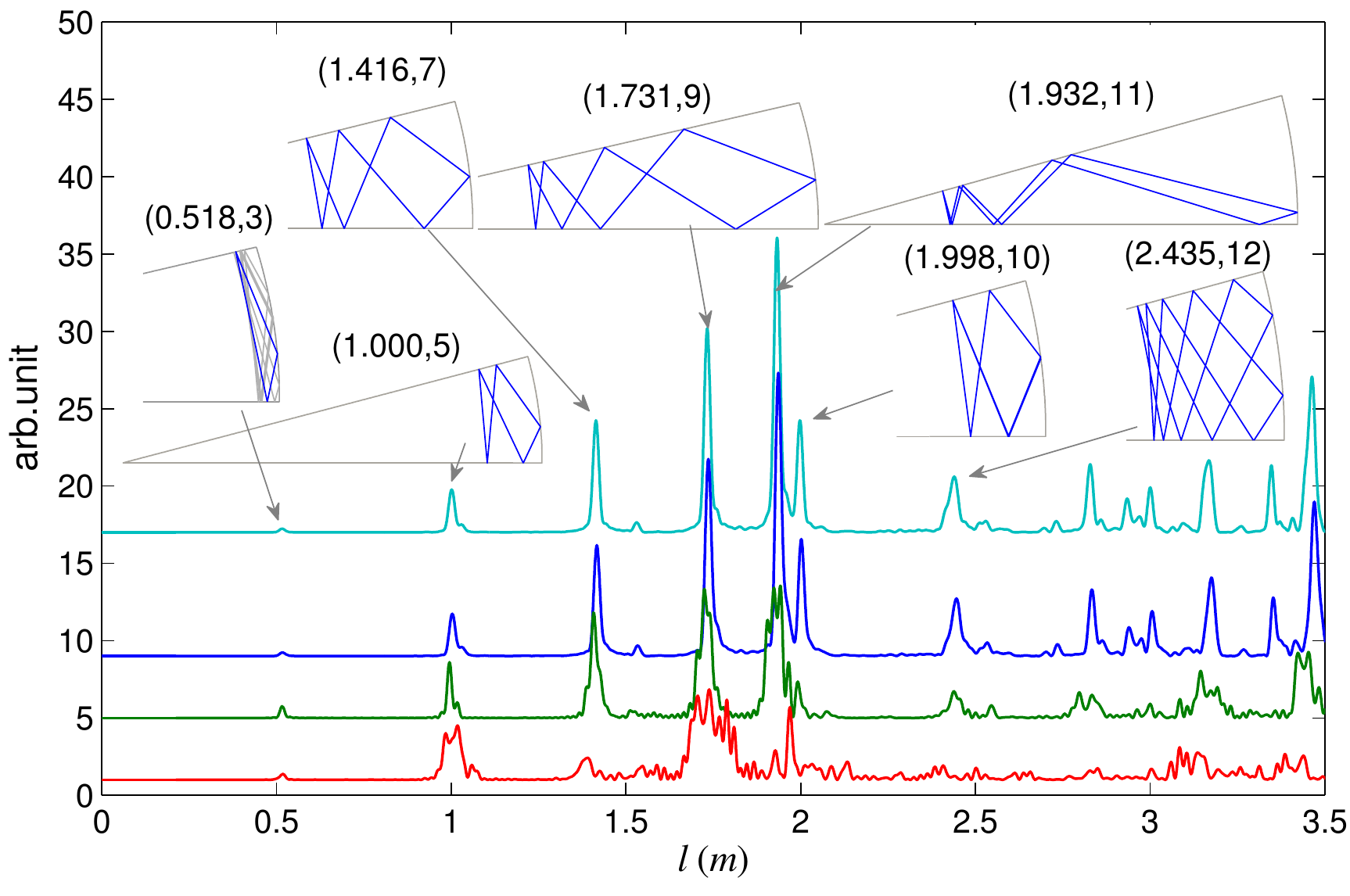,width=\linewidth}
\end{center}
\caption{(Color online) Length spectra for a billiard with the shape of a $15^{\circ}$ sector. The three lower curves show results for the graphene billiard.
From bottom to top, the curves correspond to the energy intervals $E/t\in [1.7, 2.1]$ with 14660 levels $E/t\in [2.4, 2.7]$ with 8814 levels, and
$E/t\in [2.95, 3]$ with 1509 levels.  The number of levels used for the computation of the length spectra was chosen such that the average wavevector differences equaled $\Delta k\simeq 388 m^{-1}$ to achieve comparable widths of the peaks for the three cases.
The curves for different energy ranges have been shifted with respect to each other for visualization. The topmost curve shows the length spectrum of the corresponding quantum billiard, with 1509 computed levels equaling the number of levels in the energy interval $E/t \in [2.95, 3]$, see Table~\ref{tab:sectors}. Insets:
periodic orbits with lengths corresponding to the positions of the peaks, with the two numbers in the parentheses indicating their lengths and orders.}
\label{fig:15ls}
\end{figure}

\subsection{Spectral properties of graphene sector billiards}
For illustration, we first discuss the spectral properties of a graphene billiard with the shape of a sector with angle $\pi /12$ corresponding to a $15^{\circ}$ sector. The structure of one straight boundary is purely zigzag, while the other one consists of a mixture of zigzag and armchair segments. Before analyzing the spectral properties, the energy levels $E_i$ need to be unfolded to a uniform average level density. This is done by replacing $E_i$ by the smooth part of the integrated level density,
$E^{(u)}_i = \tilde{N}(E_i)$ yielding an average spacing unity between adjacent unfolded levels. Here, $\tilde{N}(E)$ was determined from a polynomial fit to $N(E) = \int_{-\infty}^EdE^\prime\sum_i\delta(E^\prime-E_i)$. This, however, is not possible in the vicinity of the van Hove singularities at $|E/t| \approx 1$. There we accordingly used the analytical result for the level density in terms of an integral given in Ref. \cite{BK:2005}. Figure~\ref{fig:15degree1} (a) shows the density of states, Figs.~\ref{fig:15degree1} (b)-(f) the distributions of the spacings between adjacent energy levels and Fig.~\ref{fig:15degree2} the $\Delta_3$ statistics, which yields the least-squares deviation of the integrated level density of the unfolded eigenvalues from the straight line best fitting it in the interval $L$~\cite{Mehta1990}. Since classical sector billiards possess an integrable dynamics, one would expect the fluctuation properties in the spectra of the corresponding
graphene billiard to be Poisson. This is indeed the case for energy values close to the upper and the lower band edges at $E=\pm E_{edge}=\pm 3t$, as demonstrated in
Fig.~\ref{fig:15degree1} (f) and in Fig.~\ref{fig:15degree1} by the pentagons. In fact --- except for a very few states --- there is a one-to-one correspondence between the eigenwavevectors of the tight-binding Hamiltonian and the lowest ones of the corresponding quantum billiard in these energy ranges, see the inset of Fig.~\ref{fig:15degree1} (f). Similarly, the associated intensity distributions of the graphene billiard exhibit the same patterns as the squared wavefunctions of the corresponding quantum billiard, see Figs.~\ref{fig:15patterns} (a-d), again except for a few cases were we observe slight distortions, see, e.g., Fig.~\ref{fig:15patterns} (c). Actually, in these energy ranges the quasimomenta of the graphene billiard may be identified with the low-energy eigenwavevectors of the corresponding quantum billiards~\cite{Dietz2015}. Accordingly, close to the band edges, the effective wavelengths are long compared to the distances between the neighboring atoms and the lengths of the armchair and zigzag line-segments along the boundaries. Consequently, the eigenfunctions and eigenwavevectors are insensitive to small distortions of the boundary which is effectively continuous. This analogy between a graphene billiard and the quantum billiard of corresponding shape close to the band edges was observed for {\em all the sectors} considered in the present article, implying that the spectral properties coincide with Poisson statistics for $E\simeq\pm 3t$, also in the $30^{\circ}$ and $60^{\circ}$ sectors with non-perfect boundaries.

In summary, near the band edges $E/t=\pm 3$, the graphene billiards are essentially described
by the nonrelativistic Schr\"{o}dinger equation for the corresponding quantum billiard implying Poisson statistics for the fluctuations in the eigenenergy spectra. With increasing deviation of the energy values from $E/t=\pm 3$, e.g.,
for $E/t\in (2.6,2.7)$, the spectral statistics evolves toward that of random matrices from the  GOE, as exemplified in Fig.~\ref{fig:15degree1} (e) and in Fig.~\ref{fig:15degree2} by the diamonds. Some representative intensity distributions are shown in Fig.~\ref{fig:15patterns} (e-h) for energy values around $E/t=2.66$. In this energy range the structure of the boundary, which is not perfectly smooth, leads to a spectral statistics, which is not purely Poisson like in the corresponding quantum billiard with Dirichlet conditions along the whole perimeter, but intermediate between Poisson and GOE statistics.
There are some intensity distributions of which the patterns are similar to those of the squared wavefunctions of the corresponding quantum billiard close to the circular boundary. However, in distinction to the latter, they are also peaked along a circle around the tip of the sector in the interior of the billiard, such as the intensity distribution shown in Fig.~\ref{fig:15patterns} (h). As the energy is further increased, e.g., for $E/t\in [2.0,2.1]$, the spectral properties show an increasing similarity with the GOE signatures, as exemplified in Fig.~\ref{fig:15degree1} (d) and in Fig.~\ref{fig:15degree2} by the plus-signs. In this energy range, while the eigenfunctions may retain certain features of the wavefunctions of the corresponding quantum billiard, the pattern structure can also be quite different, as is visible, e.g., in the intensity distribution shown in Fig.~\ref{fig:15patterns} (e). These features are reminiscent of the fluctuation properties in the eigenvalue spectra of quantum billiards with mixed Dirichlet and Neumann boundary conditions~\cite{Sieber1995}, where in the present case the boundary conditions might not be purely Neumann or Dirichlet in some parts of the boundary. We would like to remark that mixed boundary conditions are a purely wave-dynamical feature with no classical analogue.

For energy values close to the Dirac point, excluding the edge states, the quasimomenta may be identified with the low-energy Dirac eigenvalues obtained by solving the Dirac equation for a graphene flake of corresponding lattice structure. Accordingly, like in the vicinity of the band edges, the effective wavelengths are long, so one may expect that the spatially discrete nature of the boundaries has little effect on the statistical properties of the eigenstates, and thus, the fluctuation properties in the energy spectra are described by Poisson statistics. However, firstly, the occurrence of the Dirac point has its origin in the two interpenetrating triangular sublattices forming the hexagonal lattice of graphene, that is, it is a result of the lattice structure. Secondly, our computations, as exemplified in Fig.~\ref{fig:15degree1} (b) and by the circles in Fig.~\ref{fig:15degree2}, yielded that the spectral properties agree well with those of random matrices from the GOE~\cite{ref:Diracbilliard}. While surprising, the phenomenon can be qualitatively understood from the properties of the intensity distributions. Figures~\ref{fig:15patterns}(i-l) show examples for energies close to the Dirac point, which exhibit  disordered patterns which are spread over the whole billiard plane.
Furthermore, the intensities are peaked at some of the zigzag edges of the straight boundary which is formed by a mixture of armchair and zigzag edges, like the squared wavefunctions in a quantum billiard with Neuman conditions at these parts of the boundary. The effect of the varying boundary conditions visible in the intensity patterns of the graphene billiard leads to correlations in the wave amplitudes at the sites of the corresponding sublattice and, via the hopping, between the sublattices and hence, in the present example, to level repulsion between neighboring eigenenergies. This phenomenon persists for the other graphene sector billiards. In general, we came to the result that, when the lattice structure of the graphene billiard has a higher symmetry including uniform boundary conditions for each sublattice along its perimeter, like in graphene billiards with the shapes of equilateral triangles~\cite{Rycerz:2012} or a rectangular billiard~\cite{Dietz2015}, the patterns of the intensity distributions had an ordered (periodic) structure and  spectral properties tended to be more Poisson-like. However, introducing a slight variation of the boundary conditions by removing just one row of atoms or a slight roughness by removing just a few atoms, may lead to drastic changes in the eigenstates leading to GOE statistics. This extreme sensitivity on the structure of the boundary arises but only for energies close to the Dirac point. Note, that for quantum billiards, a slight change in the shape of the boundary will not affect the nature of the spectral properties in the low-energy region. On the other hand, the introduction of a pointlike perturbation, implying a change in the boundary at a singular point, will also lead to drastic changes in the properties of the wavefunctions and the spectral properties~\cite{Seba1990,Rahav2003}. We also investigated distributions of the moduli of the wavefunction components on the lattice sites and came to the result, that they are well described by a Gaussian distribution typical for chaotic systems in energy regions where the spectral properties are described by GOE statistics. Here, we had to exclude states localized along zigzag rows like they occur near the Dirac point and the van Hove singularities~\cite{DKMR:2016}.

\subsection{Length spectra}
Figure~\ref{fig:15ls} shows length spectra~\cite{Stockmann:book,DR:2015}
for the energy ranges considered in Fig.~\ref{fig:15degree1} above $E/t=1.7$. The insets of Fig.~\ref{fig:15ls} show the periodic orbits with lengths corresponding to the peak positions. Before computing length spectra the wavevectors were normalized as described in the following. For the sector billiard, each orbit belongs to a family of periodic orbits characterized by the order and the orbital length, as demonstrated by the light gray orbits  in the inset for the shortest periodic orbit. For a graphene billiard with the shape of a sector with angle $\alpha$ and radius $L=L_0 a$, where $a=2.46${\AA} is the graphene lattice constant, we have $E=\hbar v_F q$ for energies close to the Dirac point~\cite{Wallace:1947}. Here, $q$ is the quasimomentum, which may be identified with the eigenwavevectors of the graphene billiard in the vicinity of the Dirac point, and $v_F=\sqrt{3}ta/(2\hbar)$ is the Fermi velocity. This yields $E/t=\sqrt{3}a q/2=\sqrt{3}(qL)/(2L_0)$, with $(qL)$ corresponding to the normalized, dimensionless quasimomentum. The value of $L_0$ can be obtained either from $L$ or from the total number of atoms through $L_0=\sqrt{\sqrt{3}N/(2\alpha)}$. For energies close to the band edges where~\cite{Wallace:1947} $E-E_{edge}=\pm t a^2 q^2/4$, we have $(E-E_{edge})/t=(qL)^2/(4L_0^2)$. In this energy region the quasimomenta may be identified with the eigenwavevectors of the corresponding quantum billiard. Using these relations, we determined the wavevectors in units of the radius from the calculated energy spectrum and used these to compute the length spectra and to compare them in different energy ranges. In Fig.~\ref{fig:15ls} the length spectra in the energy ranges $E/t\in [1.7,2.1]$ (first curve from bottom to top ), $E/t\in [2,4,2.7]$ (second curve), and $E/t\in [2.95,3.0]$ (third curve) are compared to the length spectrum of the corresponding quantum billiard (topmost curve). Best agreement between the latter and the length spectra of the graphene billiard is obtained for energies closest to the band edge, i.e., for $E/t\in [2.95,3.0]$, whereas the peaks broaden and additional ones become visible with decreasing energy. These observations are in line with the observed analogy between the eigenstates of the quantum and graphene billiard, which is lost with decreasing energy. For energy values close to the Dirac point we didn't find any analogy between the peak positions in the length spectra and those of the corresponding quantum billiard. This suggests that, while there can be eigenfunctions localized on periodic orbits~\cite{HLFG:2009}, the main fraction of eigenstates appear more random as compared with those close to the band edge (Fig.~\ref{fig:15patterns}).
\begin{figure}[h]
\begin{center}
\epsfig{file=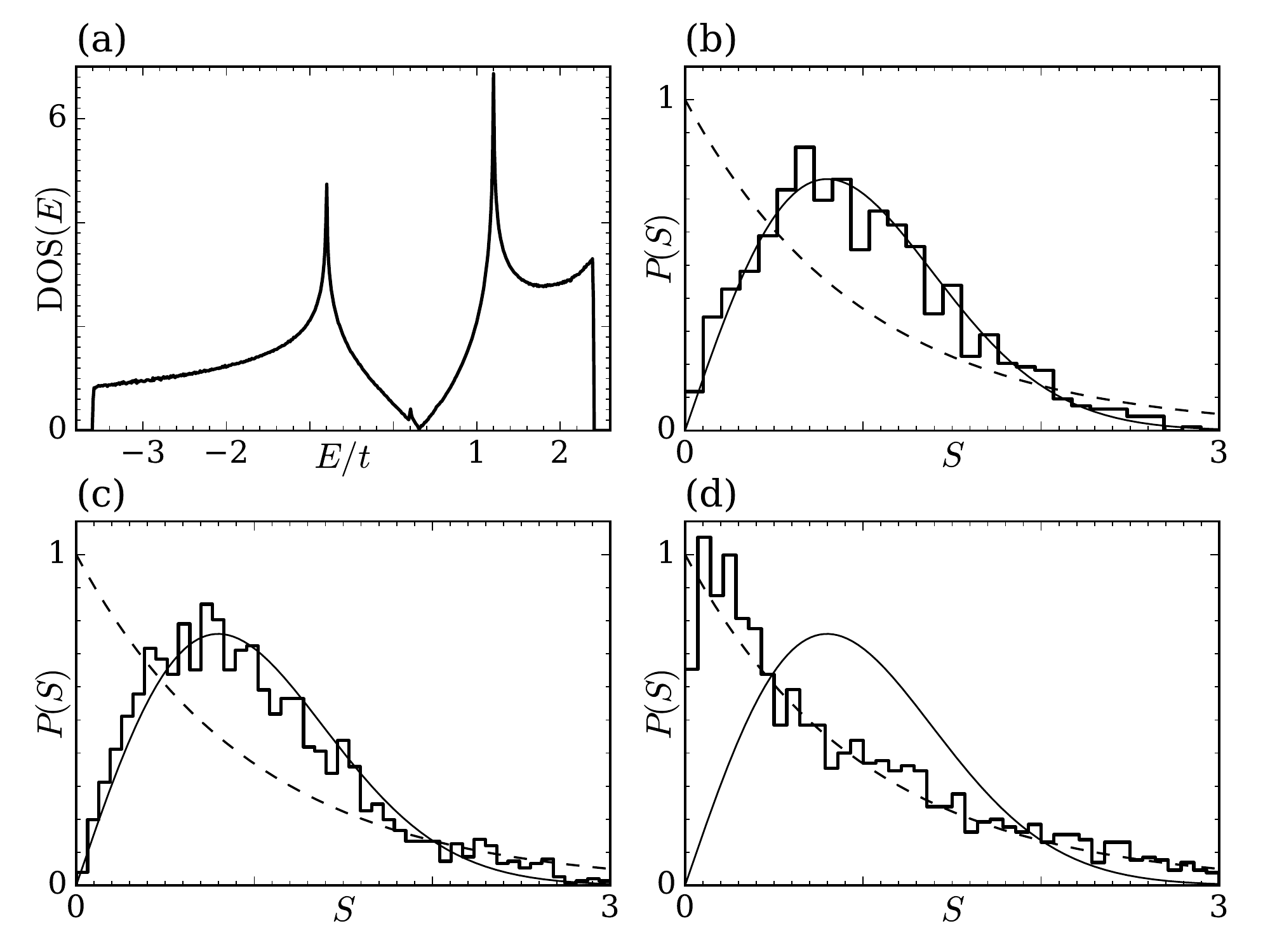,width=\linewidth}
\end{center}
\caption{Nearest-neighbor spacing distributions in the presence of next-nearest-neighbor hoppings for the $15^{\circ}$-sector graphene billiard. Choosing the next-nearest-neighbor hopping energy as $t'=0.1t$, the Dirac point is at $3t'=0.3t$.
(a) the density of states, (b-d) same as in Fig.~\ref{fig:15degree1} for
$E/t \in [0.3,0.5]$, $[-2.7,-2.6]$, and $[-3.6,-3.5]$, respectively.}
\label{fig:15degreetp}
\end{figure}

\subsection{The effect of edge states and next-nearest-neighbor interactions on the spectral properties}
A pertinent issue concerns the effects of nonvanishing next-nearest-neighbor interactions
in the graphene Hamiltonian on the spectral properties, which have been
studied previously for edge states and in chaotic graphene billiard. In particular,
in Ref.~\cite{WAG:2010} the edge states occurring close to the Dirac point were studied in a
disordered graphene billiard, with the finding that the spectral properties are of GOE type for vanishing  next-nearest-neighbor hoppings. However, due to the fact that the edge states are localized, regardless of the nature of the underlying classical dynamics~\cite{Wurmetal:2009}, in such
systems the spectral properties should be expected to be described by Poisson statistics.
The fact, that GOE statistics is observed may be attributed to the chiral symmetry of
the graphene Hamiltonian including only the nearest-neighbor hopping terms, implying that
each eigenstate occupies sublattices A and B with equal probability. Because the zigzag
segments are formed by the atoms from one sublattice, an edge state occupying one such segment
must also be nonvanishing on a zigzag segment of the boundary formed by the other sublattice.
This leads to a long-range connection between the edge states, thus explaining the observed level repulsion.
By introducing a next-nearest-hopping term $t'=0.1t$, the chiral symmetry and thus this coupling is
lifted. As a result, the spectral properties of the edge states change to Poisson statistics as they
are spatially localized.

Note, that the graphene billiards studied in Ref.~\cite{Wurmetal:2009} had shapes of billiards of which the classical dynamics was far from integrable. This may also be a contributing factor to the GOE statistics. Similar results were obtained in experiments with an Africa-shaped graphene billiard~\cite{DKMR:2016} with nonvanishing first-, second- and third-nearest-neighbor couplings of the atoms. There, the edge states had to be excluded in order to obtain pure GOE statistics. Furthermore, the spectral properties of graphene billiards with shapes of classically {\rm chaotic} billiards and nonvanishing next-nearest-neighbor couplings $t'$ were studied numerically in Ref.~\cite{HLG:2011}. There, agreement with GOE statistics was obtained, even when taking into account {\em all} energy levels close to the Dirac point. We have systematically tested the effect of nonvanishing $t'$ interactions in sector-shaped graphene billiards. A representative result for $t^\prime =0.1t$ is shown in Fig.~\ref{fig:15degreetp}, so the Dirac point is at $E=3t'=0.3$. We observe that, while the density of states differs considerably from that for $t^\prime =0t$, the spectral properties are comparable in the different energy ranges (see Fig.~\ref{fig:15degree1}), that is, in the vicinity of the Dirac point GOE statistics persists.

\subsection{Spectral properties near the van Hove singularities}
The GOE statistics found in the graphene billiards with shapes of
classically integrable sector billiards in the vicinity of the band edges and the
Dirac point were at first unexpected since the effective wavelengths are large as compared with
the distance between the neighboring atoms along the edges.
As outlined in the previous sections deviations from the expected Poisson statistics may be explained by
the mixed boundary conditions, applying to the circular and partly to the straight boundaries.
Furthermore, the condition of much larger effective wavelengths is not met particularly as the energy approaches the van Hove singularities at $E/t=1$ either downward from the upper band edge or upward from the Dirac point. There, the density of states diverges~\cite{BK:2005} logarithmically with increasing size, i.e., number of atoms forming the graphene billiard. Furthermore, the effective wavelength decreases and eventually becomes comparable to the lattice constant. In addition, the trigonal warping
effect~\cite{ref:trigonalwarping,rycerz_valley_2007,cheianov_focusing_2007,
garcia-pomar_fully_2008} becomes dominant implying that the motion of the
quasiparticle is constrained to the six angular directions
along the zigzag rows within the graphene lattice.
Note that this constraint is compatible with the lattice
symmetry, which guarantees specular reflections at the discretized
boundary. Thus, close to the van Hove singularities, the corresponding classical motion
is nonergodic. This is compatible with the features of the wavefunctions
in the vicinity of the van Hove singularities, which are localized along zigzag rows within the
hexagonal lattice~\cite{DKMR:2016}. One may therefore expect that the spectral properties
return to Poisson. In this regard, it was demonstrated experimentally and numerically on the basis of the tight-binding model that for a rectangular graphene billiard the spectral properties
at the van Hove singularity are indeed Poisson~\cite{DKMR:2016}. On the other hand, for a graphene
billiard with the shape of a chaotic Africa billiard the spectral properties were shown to coincide with GOE statistics after extraction of the energy values close to the van Hove singularities associated with wavefunctions localized along zigzag rows. However, our calculations surprisingly revealed persistent
GOE statistics even for energies near the van Hove singularity.

In order to get an idea concerning the mechanisms that lead to GOE statistics, we recall previous results for the spectral properties in nonrelativistic quantum billiards with the shapes of polygons, which support pseudo-periodic orbits in the classical limit but exhibit GOE statistics~\cite{RB:1981,CC:1989,Zyczkowski:1992, SSSS:1994}. According to Richens and Berry~\cite{RB:1981}, the GOE
statistics may be attributed to the bifurcation of the trajectories scattered at the corners of the polygons formed by the discrete boundaries. Although this picture is more accurate when the wavelength is much smaller than the distance
between the adjacent corners, corresponding to the distance between
two neighboring boundary atoms in graphene billiards,
random scattering at the boundaries can become influential when the
wavelength is comparable with the lattice constant, leading to GOE
statistics, like in rough billiards~\cite{Sirko2000}. For the energy range between the Dirac point and the van
Hove singularity, effects induced by the discretized boundary and trigonal warping are both important, leading again to GOE statistics. We have tested
that these features of the spectral properties hold for {\em all} the sectors in Table \ref{tab:sectors} with different angles. The only exception are the 60$^\circ$ sectors with perfect edges along the straight boundaries, whose statistics are the combination of two GOEs. We may conclude that in graphene billiards with the shape of classically integrable sector billiards the spectral properties are dominated by GOE statistics except in the energy range close to the band edges $E_{edge}/t \sim\pm 3$, where the quasiparticles are essentially described by the nonrelativistic Schr\"{o}dinger for quantum billiards of corresponding shape, and the fluctuation properties follow the Poisson statistics.
\begin{figure}[h]
\begin{center}
\epsfig{file=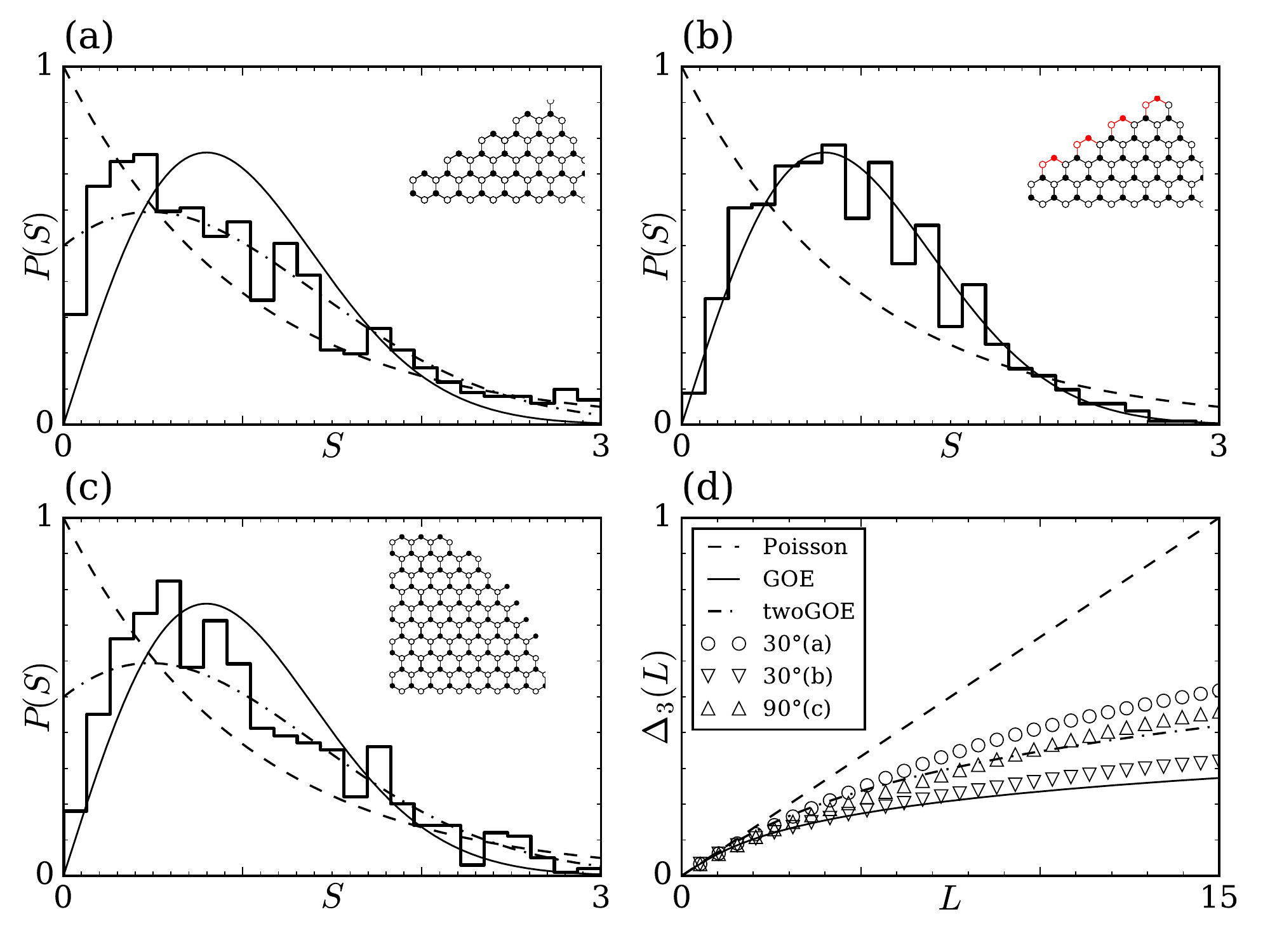,width=\linewidth}
\end{center}
\caption{(Color online) Effect of structural perturbations on the spectral properties in the energy range $E/t \in [0.02,0.2]$ near the Dirac point for a graphene billiard with the shape of a (a) $30^{\circ}$-sector, (b) a $30^{\circ}$-sector with an additional row of atoms along the armchair edge, and (c) a $90^{\circ}$ sector. (d) $\Delta_3$ statistics of the cases (a)-(c). Insets: magnified view of the lattice structure around the tip of the sector in order to illustrate the differences.}
\label{fig:oneline30}
\end{figure}
\begin{figure}[h]
\begin{center}
\epsfig{file=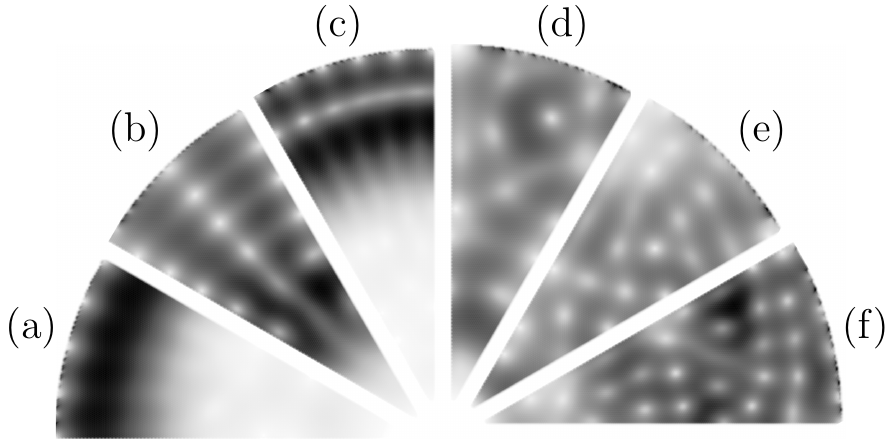,width=\linewidth}
\end{center}
\caption{Effect of structural perturbations on the intensity distributions for a graphene billiard with the shape of a $30^{\circ}$ sector. (a-c): as in Fig. \ref{fig:oneline30} (a) for energies 0.085, 0.099 and 0.117, respectively. (d-f): as in Fig.~\ref{fig:oneline30} (b) for energies 0.086, 0.099 and 0.117, respectively.}\label{fig:30patterns}
\end{figure}

\subsection{Effect of structural perturbations on the spectral properties}
When both straight boundaries of the sector coincide with zigzag edges formed by one of the two sublattices or with armchair edges, the boundary of the sector is deemed to be ordered or perfect. As a result, the spectral properties are expected to be closer to Poisson statistics as compared with those of the $15^{\circ}$ and $45^{\circ}$ sectors. In this regard, the $30^{\circ}$ and $90^{\circ}$ sector boundaries comprise one straight zigzag edge, whereas the other one
is an armchair edge; see insets of  Fig.~\ref{fig:oneline30}. Our simulations show that for all
energy ranges except that close to the Dirac point, the spectral statistics
exhibits essentially the same features as for the $15^{\circ}$ or
$45^{\circ}$ sectors where no perfect zigzag or armchair edge boundary may be realized. That is, we observe Poisson statistics near the band edges and, otherwise, GOE statistics. Close to the Dirac point, e.g., for $E/t \in [0.02,0.2]$,
the statistics is a mixture of GOE and Poisson, as shown in
Figs.~\ref{fig:oneline30} (a,c). This may be attributed to the distinct boundary conditions along the straight parts of the boundary. Under structural perturbation, e.g., when an additional row of atoms is added to the upper straight boundary
for the $30^{\circ}$ sector, indicated by the additional row of red coloured atoms in the inset of Fig.~\ref{fig:oneline30} (b), the statistics becomes GOE. These results are in accordance with those obtained in Ref.~\cite{Borgonovi1996}, where the spectral properties of a Bunimovich stadium billiard consisting of a quarter circle and an infinitesimally-sized rectangular part were studied. The same phenomenon occurs when just the four atoms at the tip of the sector are removed. Similar results were obtained, when introducing a singular perturbator into a circle billiard~\cite{Rahav2003}. Like in such a singular billiard the effect on the intensity patterns by such a change is drastic as shown in Fig.~\ref{fig:30patterns}, even though the classical orbits hitting this corner or the singular scatterer are of measure zero in classical phase space. The intensity patterns are more ordered for the $30^{\circ}$ sector with no structural perturbation.

\begin{figure}[h]
\begin{center}
\epsfig{file=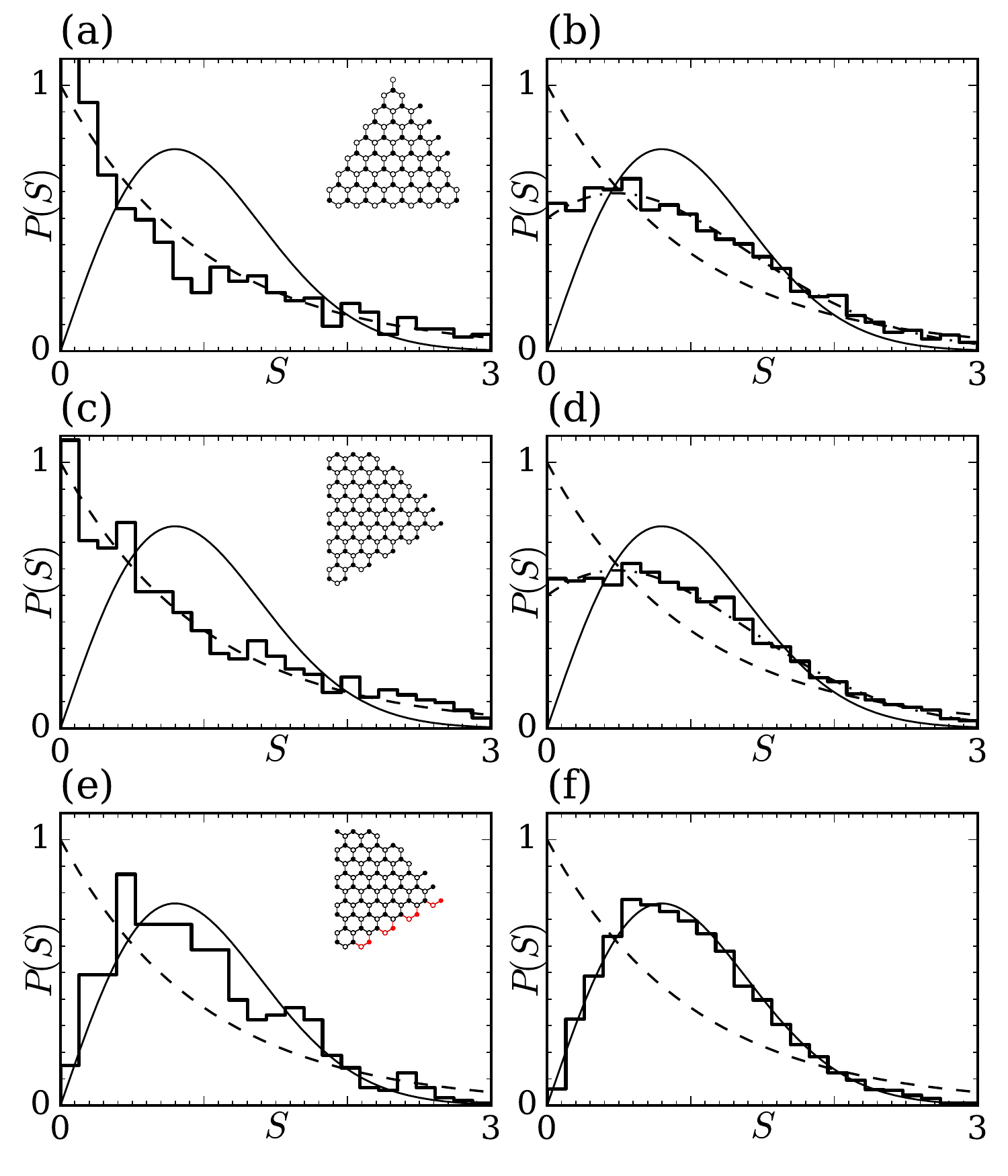,width=\linewidth}
\end{center}
\caption{(Color online) Effect of structural perturbation on the nearest-neighbor spacing distribution for a graphene billiard with the shape of a $60^{\circ}$ sector.
(a,b) zigzag edges, (c,d) armchair edges, (e,f) armchair edges
with one row of atoms removed along one edge. The left and right panels
are for $E/t\in [0.02,0.2]$ and $E/t\in [0.7,0.8]$, respectively. Insets:
magnified view of the lattice structure close to the tip of the sector to illustrate the differences.}
\label{fig:oneline60}
\end{figure}

\begin{figure}[h]
\begin{center}
\epsfig{file=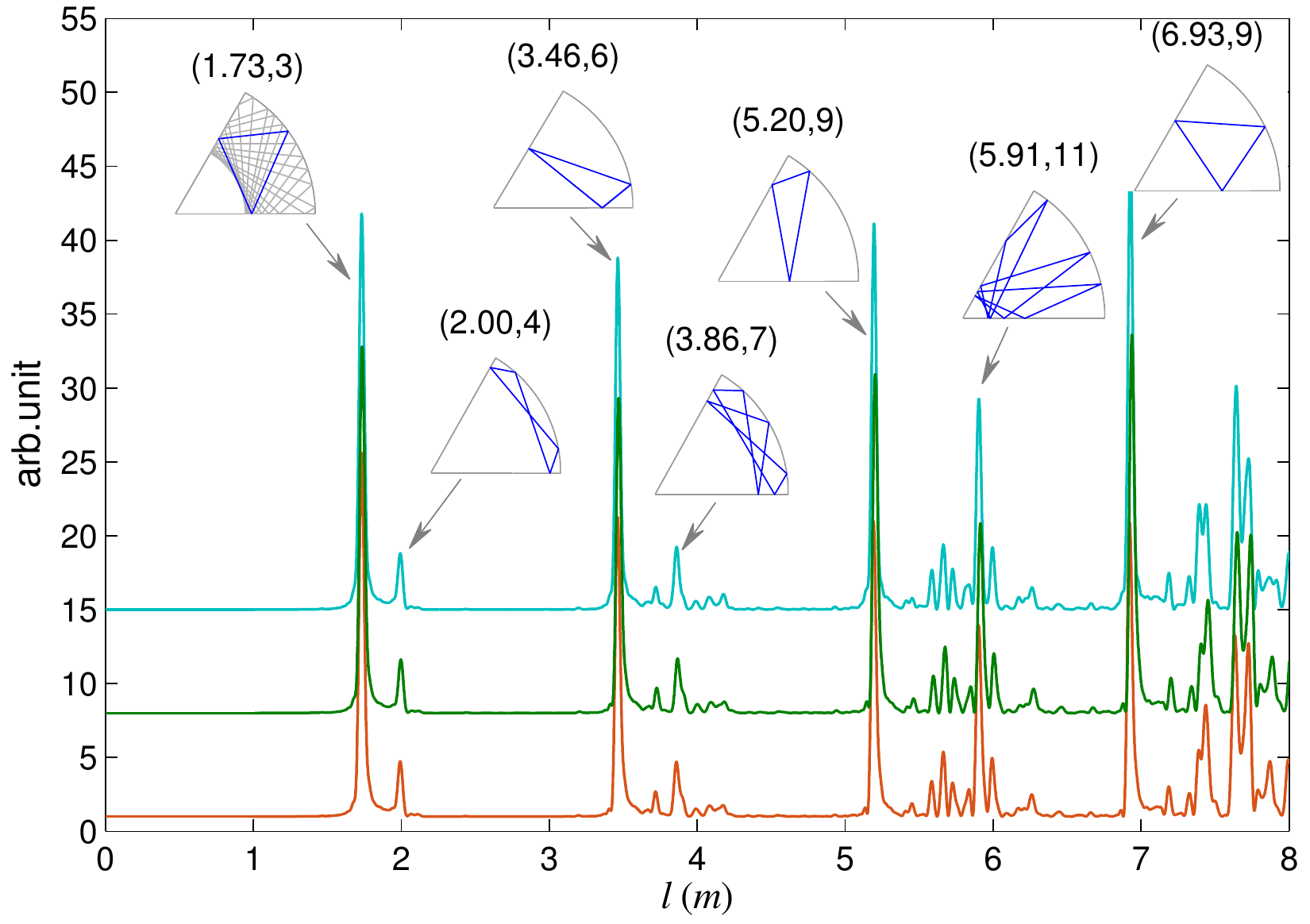,width=\linewidth}
\end{center}
\caption{(Color online) Length spectra for the $60^{\circ}$-shaped graphene billiard with armchair edges along the straight boundaries. The insets show the periodic orbits corresponding to the peaks, with the two numbers in the parentheses indicating the length and the order of the orbit, respectively. The middle and bottom curve shows the length spectrum of the graphene billiard in an energy range $E/t\in (2.95, 3)$, for the perfect case (1530 levels) and the non-perfect case (1524 levels) respectively. The topmost curve is obtained from the first 1530 eigenvalues of the quantum billiard of corresponding shape.
The different curves are shifted vertically for
clear visualization.}
\label{fig:60ls}
\end{figure}

For the $60^{\circ}$sector shaped graphene billiard, both straight parts of the boundary are zigzag edges formed by the same sublattice or armchair edges, so the system possesses an exact reflection symmetry. Alternatively,
the $60^{\circ}$ domain can be viewed as constituted by two $30^{\circ}$
sectors with differing boundary conditions along its symmetry line for symmetric and antisymmetric eigenstates, so the eigenenergies and eigenstates are related to those of the $30^{\circ}$ sector. Figure~\ref{fig:oneline60} shows the level
spacing statistics for $E/t\in [0.02,0.2]$ (left columns), and for
$E/t\in [0.7,0.8]$ (right columns). For energies close to the band edges,
the spectral statistics follow Poisson in all cases. For the other energy ranges, the nearest-neighbor spacing distributions coincide with those shown for $E/t\in [0.7,0.8]$. For either the zigzag or the armchair edge case, when the system possesses a perfect reflection symmetry, the spectral properties are close to Poisson in the vicinity of the Dirac point, in accordance with the results obtained for a rectangular graphene billiard in~\cite{Dietz2015}. However, for the zigzag edge case, a deviation from Poisson is observed close to spacing zero, $S\sim 0$. The reason is that, when the system has a perfect reflection symmetry, the eigenstates are either even or odd with respect to the symmetry line. Accordingly, near the Dirac point, the energy values corresponding to the even and odd eigenstates are nearly degenerate, the spacings being of the order of $S\sim 10^{-2}$, as compared to the average spacing unity, yielding the unusually large count in the nearest-neighbor spacing distribution observed in Fig.~\ref{fig:oneline60} (a) close to $S\sim 0$. After separation into even and odd states, this abnormally large count is removed. Also in the armchair edge case, shown in Fig. \ref{fig:oneline60} (c), there are even and odd eigenstates, but they are not nearly degenerate. Accordingly, the corresponding nearest-neighbor spacing distribution agrees with the Poisson distribution. When increasing the energy towards $E/t\in [0.7,0.8]$, the nearest-neighbor spacing distributions neither coincide with the Poisson distribution nor with the Wigner distribution, as illustrated in Fig.~\ref{fig:oneline60} (b,d), but are between Poisson and GOE. A further investigation reveals that the statistics coincides with that of two independent GOEs, i.e., they are the result of combining two sets of energy levels, each of GOE nature. The corresponding nearest-neighbor spacing distribution~\cite{Guhr1998} is shown as dash-dotted line. This is consistent with the fact that in this energy range, the spectral statistics for the $30^{\circ}$ domain is described by the GOE. This behavior persists until the energy approaches a band edge, the combination of two independent Poisson distributions still being Poisson.

Figures~\ref{fig:oneline60} (e) and (f) show the nearest-neighbor spacing distributions for $60^{\circ}$-sector shaped graphene billiards
with armchair edges subject to a structural perturbation realized by removing
one row of atoms from the right straight boundary in the inset of Fig.~\ref{fig:oneline60} (c). The perturbation breaks the reflection symmetry. Nevertheless, certain eigenstates still
exhibit features which are similar to those of the unperturbed sector billiard. Intuitively, since the structural changes are small and the boundaries retain the armchair-edge structure, one may expect that there will be little change in the statistical properties of the eigenstates. However, it turned out that the perturbation had a significant effect on the intensity distributions in that their patterns are distributed more randomly than in the unperturbed case. The effect, actually, is similar to that observed in Fig.~\ref{fig:30patterns} for the $30^{\circ}$-sector shaped graphene billiard. Also, the effect on the spectral properties is dramatic in the sense, that the nearest-neighbor spacing distributions coincide with the Wigner distribution, not with the Poisson distribution in the vicinity of the Dirac point. This agreement persists in the higher energy regime. Furthermore, it is reminiscent of the results obtained for half-circular billiards with an infinitesimal straight cut, which corresponds to omitting one row of atoms in the graphene billiard of corresponding shapes and leads to a fully chaotic classical dynamics~\cite{Ree1999}. Generally, for the $60^{\circ}$-sector shape, Poisson statistics arises only near the band edges $E/t \sim\pm 3$. A similar behavior was observed for the $15^{\circ}$ and $45^{\circ}$-sector shapes. A general feature is that, while the level spacing statistics are Poisson for graphene billiards that possess an exact lattice symmetry along the boundaries, a small structural perturbation, i.e., the removal of a few of the $2\times 10^5$ atoms forming the lattice, like the introduction of a singular perturbator into a quantum billiard with a classically regular dynamics, can induce a transition of the spectral properties from Poisson to GOE statistics, or singular statistics~\cite{Rahav2003}.

Figure~\ref{fig:60ls} shows length spectra of the $60^{\circ}$-sector shaped
graphene billiard with perfect (middle curve) and non-perfect (bottom curve) armchair edge boundaries for the energy ranges $E/t \in (2.95,3)$. For comparison, we also display the length spectrum obtained from the eigenvalues of the quantum billiard of corresponding shape (topmost curve).
We came to the result, that the removal of one row of atoms has little effect on the energy levels from the vicinity of the band edges and on the corresponding length spectrum. Furthermore,
the peaks in the length spectra for $E/t \in (2.95,3)$ agree with those
of the quantum billiard.

\section{Discussion} \label{sec:discussion}

One of the pillars in the field of quantum chaos is the discovery of
GOE characteristics in the statistics of the energy level spacing for
Hamiltonian systems that are fully chaotic in the classical
limit~\cite{BGS:1984,BR:1984,Berry:1985,CCG:1985,TM:1985,BR:1986,
DG:1986,Izrailev:1989,GKP:1991,BSS:1992,PR:1994,CS:1995,AABA:1995,
Fromholdetal:1995,JS:1997,EB:2003,HZOAA:2005,PLS:2012}. GOE statistics
has been generally regarded as an unequivocal quantum signature of classical
chaos, whereas the spectral properties of classically integrable systems
are believed to exhibit Poisson statistics~\cite{BT:1977,MT:2003,Haake:book,Stockmann:book}.
The most pronounced difference between GOE and Poisson statistics is that,
for the former, the probability for zero level spacing is essentially zero,
while for the latter this probability is maximal, that is, classical chaos
gives rise to the phenomenon of energy level repulsion in quantum systems.
The GOE fingerprint of classical chaos persists apparently into the realm
of relativistic quantum mechanics, as it has been demonstrated for
two-dimensional Dirac material systems such as graphene~\cite{PSKYHNG:2008,
Wurmetal:2009,HLG:2010,HLG:2011,HXLG:2014,DKMR:2016}. In this work we investigated the spectral properties of graphene billiards versus quantum billiards. The latter are the quantum counterpart of a classical billiard and are described by the Schr\"odinger equation with Dirichlet conditions along the boundary. It has been found, that, actually, not the shape of a billiard determines the spectral properties of the quantum system, but instead, the boundary conditions. Deviations from Poisson were found in quantum billiards with the shape of classically integrable billiard, when imposing mixed boundary conditions~\cite{Sieber1995}, or when introducing a singular perturbation~\cite{Rahav2003} or roughness along the boundary~\cite{Sirko2000}.  Note, that there is no classical counterpart to quantum billiards with mixed boundary conditions, whereas, the classical dynamics of singular billiards is integrable, because the effect of the perturbator is of measure zero, implying a singular statistics.

These observations are corroborated in the present article for graphene billiards with the shape of circular sectors with angles  $\pi/n$ with integer $n$. Classical billiards with this shape are integrable because the domain becomes a disk through $2n$ duplications of the sector. We demonstrated that the spectral properties of such graphene billiards generally coincide with GOE. Utilizing the conventional tight-binding Hamiltonian for graphene, we calculated {\em all} the eigenenergies associated with the tight-binding Hamiltonian matrix for graphene lattices of finite but large size, enabling a detailed analysis of the level spacing statistics in different energy ranges, characterized by distinct dispersion relations.

Our main results can be summarized as follows. Near
the Dirac point $E \sim 0$, the dispersion relation is linear and
the quasiparticle behaves like a massless Dirac fermion. Near the band
edge $E_{edge}\sim \pm 3t$, the dispersion relation is quadratic
and the quasiparticle is essentially described by the nonrelativistic Schr\"{o}dinger for the quantum billiard of corresponding shape.
Accordingly the classical dynamics is integrable. In the vicinity of the band edges the spectral properties exhibit Poisson statistics for all the billiard shapes studied in this article. Deviations from Poisson occur in the relativistic quantum regime close to the Dirac point $E \sim 0$. There, Poisson statistics is expected, if the boundary is formed by zigzag edges of just one sublattice or armchair edges, so the boundary conditions are uniform on both sublattices. We found that, generically, the spectral statistics are robustly GOE. In fact, there is a high degree of
sensitivity of the nature of the statistics to the symmetry of the lattice structure of the graphene billiard, which can be finely controlled through structural perturbation at the straight boundaries. Only when both straight boundaries are perfectly zigzag, formed by just one sublattice, or
armchair, the statistics was found to follow Poisson, as expected. When an arbitrarily
small perturbation is present, introduced, e.g., by removal of one row of atoms
along one straight boundary or of a few atoms at the tip of the sector, the level statistics immediately turns into GOE. For graphene billiards with non-perfect segments along the boundary like, e.g., for the $30^{\circ}$ sector or the
$90^{\circ}$ sector, in the absence of any structural perturbation the
statistics are a mixture of Poisson and GOE but an arbitrarily small
amount of perturbation leads to GOE. If the graphene billiard has
at least one non-perfect straight boundary like, e.g., for the $15^{\circ}$ or $45^{\circ}$ sectors, the statistics will be GOE even without a perturbation. The GOE statistics in the energy range close to the Dirac point can thus be attributed to the effect of the ``non-perfect'' boundaries. Our systematic computations and detailed analysis thus reveal that for graphene billiards with the shapes of circular sectors, the levels statistics is generically GOE in the relativistic quantum regime, that is, like in Schr\"odinger billiards, not the shape of the billiard, but the boundary conditions imposed on its wavefunctions, determine its spectral properties.

Thus, we may conclude that, when the lattice structure of a graphene billiard with the shape of an integrable billiard implies uniform boundary conditions for each sublattice along its perimeter, as it is the case in graphene billiards with the shapes of equilateral triangles~\cite{Rycerz:2012} or rectangles~\cite{Dietz2015}, the spectral properties coincide with those of the corresponding quantum billiard over the whole energy range. However, introducing a slight variation of the boundary conditions by removing just one row of atoms or introducing a slight roughness by removing just a few atoms, may lead to drastic changes in the spectral properties. This extreme sensitivity on the structure is generally expected to arise only for energies close to the Dirac point. We note, that choosing the hopping parameter associated with boundary atoms different from that for interior atoms of the graphene billiard or including second-nearest and third-nearest neighbor hopping did not change our results. A further interesting investigation concerning the effect of an additional Hubbard U potential in the tight-binding Hamiltonian~\cite{Ying2014} will be postponed to a forthcoming publication.

An important practical issue concerns the discretization of the system
when calculating the energy spectrum. Lattice models have been used widely
in approximating continuous dynamics as it is computationally more feasible.
However, discretization may change the behavior of the dynamics in a
significant way. For example, consider the discretization of the
Laplacian operator on a square lattice, where the originally smooth
boundaries become a high order polygon consisting of short horizontal or
vertical line segments. Assume that the classical dynamics
of a billiard with perfectly smooth boundaries is chaotic. Then, for
the discretized system with polygonal boundaries with either horizontal
or vertical line segments, the classical billiard dynamics will be at
most pseudo-integrable, ruling out chaos. In order to generate GOE
statistics, the eigenstates need to satisfy the condition that
the corresponding wavelength be much larger than the size of the line
segments on the discretized boundary. For these states, the quantum
 dynamics of scattering from the boundary would behave as if the boundary were
smooth. This requires, equivalently, that the wavevector or the energy
be small. In this case, the level spacing statistics is GOE~\cite{Haake:book}.
As the energy is increased, the wavelength becomes progressively smaller.
At a certain point, the wavelength will be comparable to the size of the
line segments on the boundary. In fact, even before this point is reached,
the dispersion relation has already been altered with a four-fold symmetry imprinted by the square lattice symmetry, and deviates from
that of the continuous system. In previous works, this energy
range was often ignored as it does not have a classical correspondence.
However, for systems such as crystals~\cite{ATFB:2016}, graphene~\cite{Novoselovetal:2004,
Bergeretal:2004,Novoselovetal:2005,ZTSK:2005,Netoetal:2009,Peres:2010,
SAHR:2011} and its photonic crystal analog~\cite{DKMR:2016}, the energy range is highly relevant.

We remark that, on a larger scale, artificial photonic~\cite{IS:2006,DR:2015}
and phononic crystals~\cite{Maldovan:2013} have also become a physical
reality for modulating electromagnetic waves or acoustic wave propagation.
Thus, not only is a small energy or frequency range that mimics the
continuous limit physically relevant, but the full energy or frequency
spectrum are accessible experimentally~\cite{Dietz2015,DKMR:2016}, the study of which can be important
for uncovering new physical phenomena or exploiting potential applications.
As demonstrated in the present article, it is necessary to investigate the fluctuations in the spectrum in the whole energy range as the quasiparticles at different energies can exhibit characteristically different behaviors.

For the graphene lattice, the dispersion relation has a six fold
symmetry, which becomes dominant as the energy is increased from the
Dirac point. Since the group velocity of the quasiparticle is proportional
to the gradient of the energy with respect to the quasimomentum, the
six fold symmetry leads to restricted motion in the zigzag row directions
~\cite{HLFG:2009} equally
dividing $2\pi$. The restricted motions conform to the mirror reflection
at the discretized boundaries as the line segments are perpendicular to
the direction of the motion. For the classically integrable dynamics such as that of sector billiards,
one may speculate that the level statistics should follow Poisson.
However, when lattice symmetries play a role in the dispersion relation,
the level statistics may follow GOE or mixed Poisson-GOE statistics. We have corroborated this result
using a square lattice that has a four-fold symmetry in a sector domain. This counterintuitive
phenomenon can be understood from the previous investigations of
quantum systems with classically pseudo-integrable dynamics, such as
polygons, where the energy levels also exhibit GOE features for high
order polygons~\cite{RB:1981,CC:1989,Zyczkowski:1992,SSSS:1994}. According
to Richens and Berry~\cite{RB:1981}, this is the consequence of sensitive
reflections at the corners of the polygons formed by the discrete boundaries.

Generally, the interplay between different types of classical dynamics and
relativistic quantum mechanics can lead to surprising phenomena. For
example, in open Hamiltonian systems such as graphene quantum dots,
previous works revealed that chaos has relatively weaker effects on the
quantum scattering dynamics as compared with the nonrelativistic quantum
counterpart~\cite{YHLG:2011,BHLG:2015}. For example, when the classical
dynamics of a quantum dot is integrable or mixed, there are sharp
transmission resonances. For a Schr\"{o}dinger particle, the resonances
are completely washed out (or significantly broadened) as the classical
dynamics becomes fully chaotic, leading to smooth variations in the
quantum transmission with the energy. However, for graphene, even for
fully developed classical chaos, there are still sharp resonances.
A similar behavior is present associated with phenomena such as chaos
regulated quantum tunneling~\cite{PLWAL:2011,LAOP:2012,NHYL:2013}, chaos
induced modulation of quantum transport~\cite{YHLP:2012,YHLP:2013}, and
superpersistent currents in Dirac fermion systems~\cite{XHLG:2015}.
Our work represents another example where the relativistic quantum
signatures of classical nonlinear dynamics can be counterintuitive.

\section*{Acknowledgement}

L.H. thanks Prof.~A. Rycerz for helpful discussions. This work was
supported by NSF of China under Grants Nos.~11135001, 11375074,
and 11422541, as well as by the Doctoral Fund of Ministry of
Education of China under Grant No.~20130211110008. Y.C.L. and H.Y.X.
were supported by AFOSR under Grant No.~FA9550-15-1-0151 and
by ONR under Grant No.~N00014-15-1-2405.

\bibliographystyle{apsrev4-1}
\bibliography{GOE}

\begin{thebibliography}{80}
\bibitem [{ref()}]{ref:Diracbilliard}%
  \BibitemOpen
  \href@noop {} {}\bibinfo {note} {We have calculated the eigenenergies of the
  Dirac equation in the same sector domain using the boundary integral method.
  They are quite different from the eigenenergies of the graphene sector
  billiard for energies close to the Dirac point.}\BibitemShut {Stop}%
\end{thebibliography}

\end{document}